 \documentclass[prl,amsmath,amssymb,twocolumn, showpacs, superscriptaddress,10pt]{revtex4-1}

\usepackage{amsmath}
\usepackage{hyperref}
\usepackage{graphicx}
\usepackage{amsfonts}
\usepackage{amsthm}
\usepackage{cases}

\usepackage{color}
\definecolor{Blue}{rgb}{0.00, 0.00, 1.00}
\definecolor{Red}{rgb}{1.00, 0.00, 0.00}

\newcommand{\be}{\begin{equation}}
\newcommand{\ee}{\end{equation}}
\newcommand{\bea}{\begin{eqnarray}}
\newcommand{\eea}{\end{eqnarray}}

\newcommand{\beq}{\begin{equation}}
\newcommand{\eeq}{\end{equation}}
\newcommand{\beqn}{\begin{eqnarray}}
\newcommand{\eeqn}{\end{eqnarray}}

\begin{document}

\title{Exact short-time height distribution in 1D KPZ equation and
edge fermions at high temperature}

\author{Pierre Le Doussal}
\affiliation{CNRS-Laboratoire de Physique Th\'eorique de l'Ecole Normale Sup\'erieure, 24 rue Lhomond, 75231 Paris Cedex, France}
\author{Satya N. \surname{Majumdar}}
\affiliation{LPTMS, CNRS, Univ. Paris-Sud, UniversitŽ Paris-Saclay, 91405 Orsay, France}
\author{Alberto \surname{Rosso}}
\affiliation{LPTMS, CNRS, Univ. Paris-Sud, UniversitŽ Paris-Saclay, 91405 Orsay, France}
\author{Gr\'egory \surname{Schehr}}
\affiliation{LPTMS, CNRS, Univ. Paris-Sud, UniversitŽ Paris-Saclay, 91405 Orsay, France}

\date{\today}

\begin{abstract}
We consider the early time regime of the Kardar-Parisi-Zhang (KPZ) equation in $1+1$ dimensions in curved (or droplet)
geometry. We show that for short time $t$, the probability distribution $P(H,t)$ of the height $H$ at a given point $x$ takes
the scaling form $P(H,t) \sim \exp{\left(-\Phi_{\rm drop}(H)/\sqrt{t} \right)}$ where the rate function $\Phi_{\rm drop}(H)$ is computed exactly. 
While it is Gaussian in the center, {\it i.e.}, for small $H$, the PDF has highly asymmetric non-Gaussian tails which we characterize in detail. This function $\Phi_{\rm drop}(H)$ is surprisingly reminiscent of the large deviation function describing the stationary fluctuations of finite size models belonging to the KPZ universality class. Thanks to 
a recently discovered connection between KPZ and free fermions, our results have interesting implications for the fluctuations of the rightmost fermion in a harmonic trap at high temperature
and the full couting statistics at the edge. 
\end{abstract}

\pacs{05.40.-a, 02.10.Yn, 02.50.-r}


\maketitle

It is by now well known that many stochastic growth models in one dimension
belong to the celebrated Kardar-Parisi-Zhang (KPZ) universality 
class \cite{KPZ,directedpoly,reviewCorwin}. These models are usually described by a field
$h(x,t)$ that denotes the height of a growing interface at point $x$ at time $t$.
At the center of this class resides the continuum KPZ equation \cite{KPZ} 
where the height evolves as 
\be
\label{eq:KPZ}
\partial_t h = \nu \, \partial_x^2 h + \frac{\lambda_0}{2}\, (\partial_x h)^2 + \sqrt{D} \, \xi(x,t) \;,
\ee
where 
$\xi(x,t)$ is a Gaussian white noise with zero mean and 
$\langle \xi(x,t) \xi(x',t')\rangle = \delta(x-x')\delta(t-t')$. 
We use everywhere
the natural units of space $x^*=(2 \nu)^3/(D \lambda_0^2)$, 
time $t^*=2(2 \nu)^5/(D^2 \lambda_0^4)$ and height
$h^*=\frac{2 \nu}{\lambda_0}$.
At late times in all these growth models, including the KPZ
equation itself, while the average height increases linearly
with $t$, the typical fluctuations around the mean height 
grow as $\sim t^{1/3}$ \cite{directedpoly}. Moreover even the probability distribution function
(PDF) of the centered and scaled height is universal
and is described by the Tracy-Widom (TW) \cite{TWAll} and Baik-Rains \cite{baik2000,spohn2000} 
distributions, with a parameter that depends on the class of initial conditions
(flat, droplet, stationary)
~\cite{johansson,baik2000,spohn2000,SS10,CLR10,DOT10,ACQ11,reviewCorwin}.
Some of these predictions have also been
verified in experiments \cite{takeuchi,myllys,HH-TakeuchiReview}. 

Recently it was shown \cite{LargeDevUs} that these models undergo a third order phase transition
at late times from a strong to weak coupling phase \cite{rmt_review, colomo}. The signature of this
transition is captured by the large deviation rate functions that characterize 
atypical fluctuations of the height of order $t$. For example for the
continuum KPZ equation for the droplet initial condition, the distribution $P(H,t)$
of the height $H$ at a given space point (suitably centered) 
takes the form at large times \cite{LargeDevUs}
\begin{numcases}{\hspace*{-0.5cm}P(H,t=) \sim}
 e^{- t^2 \Phi_-(H/t)}\,,\, \Phi_-(z) = \frac{|z|^3}{12} \,, \,\, z<0
\label{largeleft}  \\
e^{- t \Phi_+(H/t) } \,,\,\,\, \Phi_+(z) = \frac{4}{3} z^{3/2} \,,\,z>0 \label{largeright}
\end{numcases}
while the central region $H \sim t^{1/3}$ is governed by the TW distribution
associated with the Gaussian Unitary Ensemble (GUE). The result in the
right tail in (\ref{largeright}), also holds for the flat initial condition. 


\begin{figure}  
\begin{center}
\includegraphics[width = 0.8\linewidth]{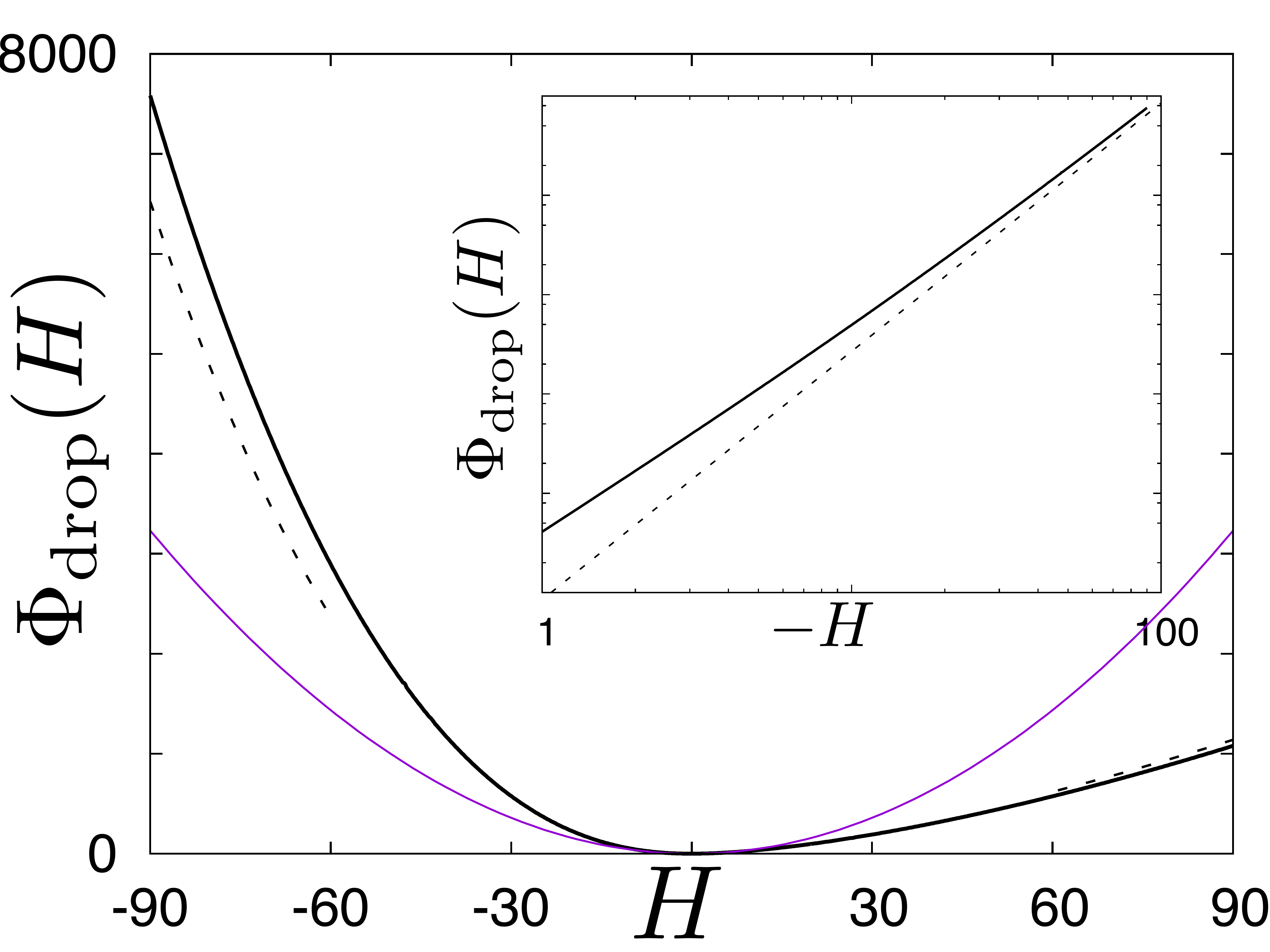}
\caption{The rate function $\Phi_{\text{drop}}(H)$ (solid black line) 
which describes the distribution (\ref{result1}) of the KPZ height $H=H(t)$ at small time 
obtained in (\ref{PhiResult}). The dashed black lines correspond respectively to the left and right tails
given in (\ref{asympt1}), (\ref{asympt3}) and the purple line corresponds to the Edward-Wilkinson
Gaussian regime (\ref{asympt2}). {\bf Inset}: Log-log plot of the left tail compared 
with the asymptotics (\ref{asympt1}).}
\label{fig:PHIdrop}
\end{center}
\end{figure}
It is then natural to wonder: are these tails $P(H,t) \sim e^{- |H|^3/(12 t)}$ ($H$ large negative) and
$\sim e^{- \frac{4}{3} H^{3/2}/t^{1/2}}$ ($H$ large positive) visible only at late times, or
do they appear even at early times? It is well known that the
central typical regime at early times is described by a Gaussian
-- obtained from the Edwards-Wilkinson's (EW) equation \cite{EW} 
setting $\lambda_0=0$ in the KPZ equation. 
What about the tails? Recently, Meerson et al. \cite{Baruch} studied this
question for the flat initial condition using the weak noise theory (WNT)
(see also earlier results \cite{Korshunov}), valid for 
short times. For the right
tail they found $\sim e^{- \frac{4}{3} H^{3/2}/t^{1/2}}$, i.e. the same leading order result as 
late times. This shows that the asymptotic right tail is established 
even at early times. In contrast, for the left tail they found 
$P(H,t) \sim e^{- \frac{8}{15 \pi} |H|^{5/2}/t^{1/2}}$ at early times
(in our units). This $|H|^{5/2}$ tail behavior for large negative $H$ for flat initial
condition is manifestly different from the $|H|^3$ tail behavior at late
times for the droplet initial condition. This raises the question 
whether this {difference} 
is due to the change in initial conditions,
or whether early and late time left tails are different for any given initial condition. 

In this Letter, we show that the early time PDF $P(H,t)$ for the
droplet initial condition takes the form
\be \label{result1}
P(H,t) \sim \exp\left( - \frac{ \Phi_{{\rm drop}}(H) }{\sqrt{t}}\right) , \quad H ~ \text{fixed} ~ \text{and} ~ t \ll 1
\ee
where $\Phi_{{\rm drop}}(H)$ is given explicitly by Eq. (\ref{PhiResult}) below. The asymptotic behaviors
of $\Phi_{{\rm drop}}(H)$ are obtained 
{as}
\begin{numcases}{\hspace*{-0.5cm}\Phi_{\rm drop}(H) \simeq}
\frac{4}{15 \pi} |H|^{5/2} \quad , \quad H \to - \infty  \label{asympt1} \\
\frac{H^2}{\sqrt{2 \pi}} \quad , \quad |H| \ll 1 \label{asympt2} \\
\frac{4}{3} H^{3/2} \quad , \quad H \to + \infty \;.
\label{asympt3} 
\end{numcases}
The first three cumulants of $H$ obtained from Eqs. (\ref{result1}) and (\ref{PhiResult}) are in agreement with 
the leading small time behavior obtained in \cite{CLR10}, while here
we obtain all cumulants. In the case of the flat initial condition
one expects a similar form \cite{Baruch},
$P(H,t) \sim e^{- \frac{ \Phi_{{\rm flat}}(H) }{\sqrt{t}} }$. However, 
the function $\Phi_{{\rm flat}}(H)$ has not been
obtained explicitly apart from the tails \cite{footnote1,footnoteGaussian} 
and the first two cumulants 
\cite{Baruch,flatshorttime}. Therefore the $|H|^{5/2}$
left tail seems to hold for a variety of initial conditions. 
Interestingly, as discussed below, our main results, Eqs. (\ref{result1}) and (\ref{PhiResult}),
turn out to be very reminiscent of the universal large deviation fluctuations
obtained in the stationary regime of finite-size models in
the KPZ universality class \cite{DerridaLebowitz,DerridaAppert,bb-00,Povo,LeeShortTime,CurrentASEP,Prolhac_PhD}.

Remarkably, these results for the 1D {\it classical} KPZ equation can be applied 
to an apriori different {\it quantum} problem 
of $N$ non-interacting fermions in a one-dimensional harmonic trap, using a 
{recent}
mapping between the two problems \cite{FermionsUS}. 
Under this mapping, the time $t$ in the KPZ equation
corresponds to $N/T^3$ where $T$ is the (dimensionless) temperature of
the fermionic system. In particular it was shown \cite{FermionsUS} that the fluctuations of the 
(dimensionless) position of the {\it rightmost fermion} ${\sf x}_{\rm max}$ near the edge ${\sf x}_{{\rm edge}}$ of the Fermi gas,
{are} 
related to
those of the KPZ height at the origin $h(0,t)$, {as} 
\bea
\frac{{\sf x}_{{\rm max}}-{\sf x}_{{\rm edge}}}{w_N} \equiv_{\rm in \, law}
\frac{ h(0,t) + \frac{t}{12} + \gamma}{t^{1/3}} 
\eea 
where $\equiv_{\rm in \, law}$ means identical PDF's. Here, on the l.h.s. $N$ is large, ${\sf x}_{{\rm edge}} = \sqrt{2N}$ and 
$w_N=N^{-1/6}/\sqrt{2}$. On the r.h.s. $\gamma$ is a Gumbel distributed
random variable with PDF given by $P(\gamma)=e^{- \gamma - e^{- \gamma}}$, independent
of the height $h(0,t)$. This equivalence in law is valid in the limit
of $N \to +\infty$, $T \to +\infty$ but with the ratio $t = N/T^{3}$
fixed. Therefore our short time results for the KPZ equation,
lead to exact predictions (\ref{fermion1}) for the high temperature behavior 
of the rightmost fermion.

Here for definiteness we focus on the narrow wedge initial condition, 
$h(x,0) = - |x|/\delta - \ln(2 \delta)$, with $\delta \ll 1$. 
This initial condition gives rise to a curved (or {\em droplet}) 
mean profile as time evolves
~\cite{SS10,CLR10,DOT10,ACQ11,reviewCorwin}.
We focus on the shifted height at a given space point, and define
\cite{footnote2}
\bea \label{H}
H(t) = h(x,t) + \frac{x^2}{4 t} + \frac{t}{12} + \frac{1}{2} \ln(4 \pi t) \;.
\eea 
The starting point of our calculation is the exact formula 
for the following generating function, obtained in
~\cite{SS10,CLR10,DOT10,ACQ11}
\begin{eqnarray}
&&\bigg\langle \exp \left( - \frac{e^{H(t) - s t^{1/3}}}{\sqrt{4 \pi t}}  \right) \bigg\rangle = Q_t(s) \label{eq:exact1} \\
&&Q_t(s) := {\rm Det}[ I - P_0 K_{t,s} P_0] \label{FD1} 
\end{eqnarray}
where $\langle \ldots \rangle$ denotes an average over the KPZ noise. Here 
$Q_t(s)$ is a Fredholm determinant associated to
the kernel
\be \label{Kt} 
 K_{t,s}(r,r') := \int_{-\infty}^{+\infty} du \, Ai(r+u) Ai(r'+u) \sigma_{t,s}(u) 
\ee
defined in terms of the Airy function $Ai(x)$ and the weight functions
\bea \label{sig} 
&& \sigma_{t,s}(u) := \sigma(t^{1/3}(u-s)) \quad , \quad \sigma(v) := \frac{1}{1+e^{- v } } \;.
\eea
In (\ref{FD1}), $P_0$ denotes the projector on the interval $r \in [0,+\infty[$. In principle,
the formula (\ref{eq:exact1}) allows us to obtain, via a Laplace
inversion, the PDF of $H(t)$ for arbitrary $t$.
The resulting expression \cite{SS10,CLR10,ACQ11} is
quite complicated: it has been analyzed at large time, but is not
very convenient for a finite time analysis. We now 
show how to extract the small time behavior directly from 
the generating function (\ref{eq:exact1}). 

It is convenient to introduce the kernel
\bea \label{defKbar} 
\bar K_{t,s}(u,u') = K_{\rm Ai}(u,u') \sigma_{t,s}(u') 
\eea 
defined in terms of the {Airy kernel 
\be
K_{\rm Ai}(u,u') = \int_0^{+\infty} dr Ai(r+u) Ai(r+u') \label{defAiry1}  
\ee
}From (\ref{defKbar}) and (\ref{defAiry1}), one checks that 
${\rm Tr}\, \bar K_{t,s}^p={\rm Tr} (P_0 K_{t,s} P_0)^p$
for any integer $p \geq 1$, which allows us to rewrite \cite{footnote_Fredholm}
\be
\! \ln {\rm Det}[ I - P_0 K_{t,s} P_0] = \ln {\rm Det}[ I - \bar K_{t,s} ] = \sum_{p=1}^{+\infty} \frac{-1}{p} {\rm Tr}\,  \bar K_{t,s}^p \label{sump} 
\ee
a convenient form to study the small $t$ limit.

We now illustrate the small time analysis on the first term $p=1$ 
of this series, the general term being analyzed in \cite{SuppMat}. 
One has
\bea
&& {\rm Tr}\, \bar K_{t,s}  = \int_{-\infty}^{+\infty} du K_{\rm Ai}(u,u) \sigma(t^{1/3}(u-s)) \nonumber \\
&& = t^{-1/3} \int_{-\infty}^{+\infty} dv K_{\rm Ai}\left(\frac{v}{t^{1/3}} , \frac{v}{t^{1/3}} \right)  \sigma(v - \tilde s) 
\label{Tr1} 
\eea 
where we have performed the change of variable $u= v/t^{1/3}$ and defined
$\tilde s=s t^{1/3}$. We see on this equation that the small $t$ limit is controled by
the large argument behavior of the Airy kernel. Since it is decreasing exponentially
fast at positive large arguments, we only need its behavior for large negative arguments.
To treat arbitrary $p$ in the equation (\ref{sump}) we need the following 
asymptotic estimate, valid for $v<0$ and $w$ fixed (see \cite{SuppMat})
\bea
K_{\rm Ai}\left(\frac{v}{t^{1/3}} ,\frac{v + t^{1/2} w}{t^{1/3}}\right)  \simeq_{t \ll 1} 
\frac{ 1}{\pi t^{1/6}} \frac{\sin \sqrt{|v|} w}{w} \,. \label{estimate} 
\eea 
We can thus replace $K_{\rm Ai}(\frac{v}{t^{1/3}} , \frac{v}{t^{1/3}})$
by $\frac{\sqrt{|v|}}{\pi t^{1/6}}  \theta(-v)$ in (\ref{Tr1}).
This leads to leading order for small $t$
\bea\label{small_time_estim}
{\rm Tr}\, \bar K_{t,s}  \simeq \frac{1}{\sqrt{t}} I_1(\tilde s)  \;,
\eea
where we 
defined 
$I_p(\tilde s) := \frac{1}{\pi} \int_{-\infty}^0 dv \sqrt{|v|} \left( \sigma(v-\tilde s) \right)^p$. Remarkably, this small $t$ estimate (\ref{small_time_estim}) can be generalized to all $p$ to obtain
the leading behavior ${\rm Tr}\, \bar K_{t,s}^p  \simeq I_p(\tilde s)/\sqrt{t}$ \cite{SuppMat}.
The series (\ref{sump}) can then be summed up, leading to
\be \label{resQ} 
\ln Q_{t}(s) \simeq - \frac{1}{\sqrt{t}} \Psi(e^{-\tilde s}) \;, \; \Psi(z) = - \frac{1}{\sqrt{4 \pi}} Li_{\frac{5}{2}}(-z) 
\ee 
in terms of the poly-logarithm function $Li_{\nu}(x) = \sum_{p=1}^{+\infty} {x^p}/{p^\nu}$. 

Hence the exact formula for the generating function (\ref{eq:exact1}) 
takes the following form at small time
\bea \label{genfunct} 
\bigg \langle \exp\left( - \frac{z}{\sqrt{4 \pi t}} e^{H(t)}  \right) \bigg \rangle \sim e^{ -  \frac{1}{\sqrt{ t}} \Psi(z) } 
\eea 
where we use $z=e^{-\tilde s}$. Note that the l.h.s. is finite only for $z>0$ (for
$z<0$ it is infinite). 

From this, assuming the
form (\ref{result1}) and inserting it in (\ref{genfunct})
for any $z>0$, we obtain $\Phi_{\rm drop}(H)$ by a saddle point analysis
\cite{SuppMat}, as
\bea
\! \! \Phi_{\rm drop}(H) = \begin{cases}
 \frac{-1}{\sqrt{4 \pi}} \displaystyle \min_{\substack{z \in [-1,+\infty[}} [ z e^{H} +  Li_{\frac{5}{2}}(-z) ], \; H \leq H_c\\
\\
 \frac{-1}{\sqrt{4 \pi}} \displaystyle \min_{z \in [-1,0[} [ z e^{H} +  Li_{\frac{5}{2}}(-z) \label{PhiResult} \\
~~~~~~~~~~~~~~~  - \frac{8 \sqrt{\pi}}{3} (- \ln (-z))^{\frac{3}{2}} ], \; H \geq H_c
\end{cases} 
\eea 
where $H_c = \ln \zeta(3/2) = 0.96026..$. Note that despite the two apparent branches, the function
$\Phi_{\rm drop}(H)$ is analytic at $H=H_c$. 
From this expression one obtains the asymptotic behaviors 
given in Eqs. (\ref{asympt1}-\ref{asympt3})~\cite{SuppMat}. One can also compute 
the cumulants of the height as, $\overline{H(t)^q}^c = t^{\frac{q-1}{2}} \phi^{(q)}(0)$, where
$\phi^{(q)}$ is the $q$-th derivative of 
\be\label{phip1}
\phi(p) = \max_H (p H - \Phi_{\rm drop}(H))  \, .
\ee
We display here the first four cumulants
\bea
&& \overline{H^2}^c = \sqrt{\frac{\pi}{2}} t^{1/2} \quad , \quad 
\overline{H^3}^c = (\frac{8}{3 \sqrt{3}} - \frac{3}{2}) \pi t \\
&& \overline{H^4}^c = (18+15\sqrt{2}-16 \sqrt{6}) \frac{\pi^{3/2}}{3} t^{3/2} \;.
\eea
Remarkably, these cumulants are very similar to
the ones obtained for the {\it stationary} fluctuations of the total integrated particle current 
for the TASEP on a finite ring \cite{DerridaLebowitz,DerridaAppert}.
This similarity holds for all higher cumulants as well (see \cite{SuppMat}). 
In fact, the generating function associated with these cumulants, called $G$ in
\cite{DerridaLebowitz,DerridaAppert} also appears in the stationary
regime of the ASEP and of the KPZ equation on a finite ring 
\cite{CurrentASEP,Povo,bb-00,LeeShortTime,Prolhac_PhD}, and is different, but similar to
our function $\Psi$. It remains a puzzle why this
universal function $G$ describing the late time stationary regime in a {\em finite} system
should be similar to our short time large deviation function in an {\em infinite} system.

\begin{figure}  
\begin{center}
\includegraphics[width = \linewidth]{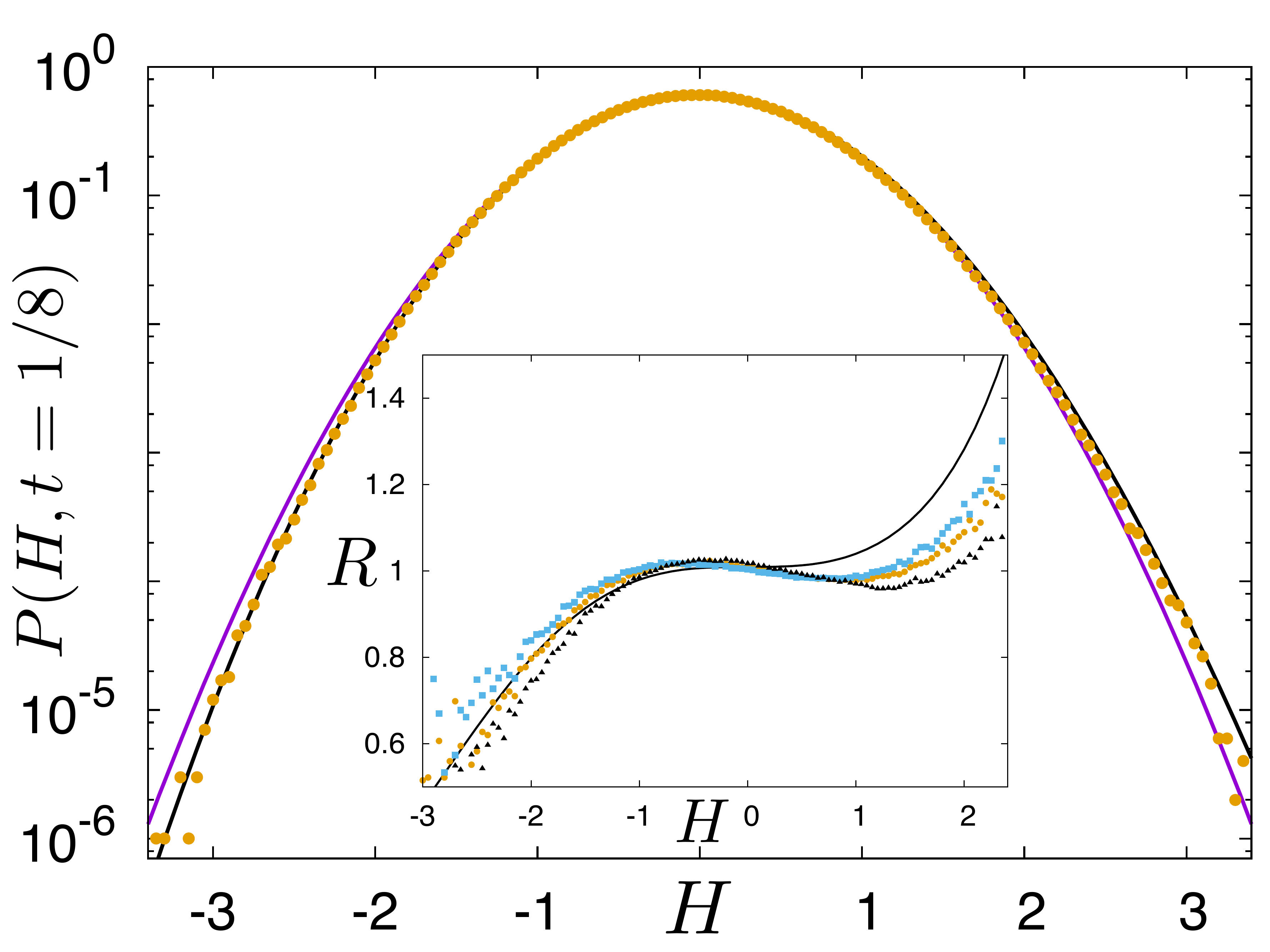}
\caption{Numerical determination of $P(H,t = 1/8)$. In the main figure, the purple line
corresponds to the Edward-Wilkinson Gaussian regime (\ref{asympt2}). The symbols correspond
to the numerical data for the discrete model with $\hat t =256, \beta=1/16$ (with $2 \,. 10^8$ samples). Note
that we have imposed $\langle H \rangle = 0$. The black line corresponds to
$P(H,t) = c(t) e^{-\Phi_{\text{drop}}(H)/\sqrt{t}}$ (\ref{PhiResult}) where $c(t) = \int_0^\infty dH e^{-\Phi_{\text{drop}}(H)/\sqrt{t}}$, with $c(1/8) = 1.23487\ldots$.
{\bf Inset:} Plot of the ratio $R = P(H,t)/P_{\rm Gauss}(H,t)$ for $t=1/8$, where $P_{\rm Gauss}(H,t)$ corresponds to the Gaussian regime (\ref{asympt2}). The triangles, circles and squares correspond respectively to $\hat t = 128$, $\hat t = 256$ and $\hat t = 512$. We see that when $\hat t$ increases the agreement with the short time continuum limit improves.}
\label{fig:PHIdrop2}
\end{center}
\end{figure}

{Our results also describe the high temperature limit of 
lattice directed polymer models (DP), which allows
for a numerical test. We simulate a DP growing on a 2D square lattice
with unit Gaussian site disorder.
For inverse temperature $\beta \ll 1$, the number of steps $\hat t$ 
corresponds to the time of the continuum model as
$t=2 \hat t \beta^4$ \cite{CLR10} (see \cite{SuppMat} for details).
The result for $P(H,t)$ is shown in Fig. \ref{fig:PHIdrop2}: the data
shows (slow) convergence to our prediction.}

{\it Fermions in an harmonic trap}. Consider now the quantum problem of
$N$ non-interacting spinless fermions of mass $m$ in an harmonic trap
at finite temperature $T$, described by the 
Hamiltonian $H=\sum_{i=1}^N \frac{p_i^2}{2 m} + \frac{1}{2} m \omega^2 {x}_i^2$. 
We use ${\sf x}^*=\sqrt{\hbar/m \omega}$ and $T^*=\hbar \omega$ as units
of length and energy. At $T=0$, i.e. in the ground state, and for large $N$, the average 
fermion density is given
by the Wigner semi-circle law, with a finite support $[-{\sf x}_{\rm edge},{\sf x}_{\rm edge}]$
where ${\sf x}_{\rm edge}=\sqrt{2 N}$. At finite temperature, the behavior of physical quantities
in the bulk changes on a temperature scale $T \sim N$ (bulk scaling), while near the edge it
varies on a scale $T = N^{1/3}/b$ (edge scaling),
where $b$ is a dimensionless parameter of order unity~\cite{FermionsUS}. 
Here we are interested in
the position ${\sf x}_{\rm max}(T)$ of the rightmost fermion (see
\cite{FermionsUS} for a precise definition). Its
cumulative distribution function (CDF)
was shown~\cite{FermionsUS} to be given by
the {\it same} Fredholm determinant as in Eq. (\ref{FD1})
\bea
{\rm Prob}\left(\frac{{\sf x}_{max}(T) - {\sf x}_{\rm edge}}{w_N} < s \right) = Q_{t=b^3}(s) 
\label{xmax_fred.1}
\eea
where $w_N=N^{-1/6}/\sqrt{2}$.
Since we have already analysed the small time limit $t\ll 1$ of the Fredholm determinant $Q_t(s)$
(as in Eq. (\ref{resQ})), this provides us with  
an explicit formula for 
the fermion problem, valid in the 
high temperature region $b \ll 1$ of the edge scaling regime. 

To use the result in (\ref{resQ}), we first set $s= {\tilde s}/t^{1/3}$ in (\ref{xmax_fred.1}) 
where $t^{1/3}=N^{1/3}/T$. The regime $t\ll 1$ corresponds to $T\gg N^{1/3}$.
This leads us to define
a new random variable
\be \label{xi}
\xi = \frac{{\sf x}_{\max}(T) - {\sf x}_{\rm edge}}{w_N(T)}
\ee
with
\bea
\label{wNT}
w_N(T) := T N^{-1/3} w_N = T/\sqrt{2 N}
\eea
where $w_N=N^{-1/6}/\sqrt{2}$ is the scale of fluctuations of ${\sf x}_{\rm max}$ at $T=0$. 
Thus $w_N(T)$ in (\ref{wNT}) sets the
scale of fluctuations of ${\sf x}_{\rm max}$ for $T\gg N^{1/3}$. 
Using (\ref{resQ}) in the limit 
$t=b^3=N/T^3 \ll 1$, we find that the CDF of $\xi$
takes the asymptotic form (replacing $\tilde s$ by $s$ for convenience) 
\be \label{fermion1} 
{\rm Prob}( \xi < s )  \sim  \exp\left( \sqrt{\frac{T^3}{4 \pi N}} Li_{5/2}( - e^{- s })  \right)\, . 
\ee 
Using $Li_{5/2}(y) \simeq y$ for small $y$, it is 
easy to see that the PDF of $\xi$ is peaked around 
the typical value $\xi=\xi_{\rm typ} = \frac{1}{2}  \ln(T^3/4 \pi N)$, with
typical fluctuations $\tilde \xi =\xi-\xi_{\rm typ}$ described by a Gumbel law, 
i.e., $P(\tilde \xi)=e^{-\tilde \xi-e^{-\tilde \xi}}$ (see \cite{Joh07} for a similar
observation in a related model). 

Our formula (\ref{fermion1}), however,
holds beyond the typical fluctuation regime and also describes
the large deviations away from $\xi_{\rm typ}$. While the
right tail is exponential, as given by the Gumbel distribution, 
using $Li_{\frac{5}{2}}(-z) \simeq_{z \to +\infty} - \frac{8}{15  \sqrt{\pi}} (\ln z)^{\frac{5}{2}}$
in (\ref{fermion1}), we find that the
left tail exhibits a distinct, stretched exponential decay
\be
{\rm Prob}( \xi < s ) 
\sim \exp\left( -  \frac{4}{15 \pi} \sqrt{\frac{T^3}{N}} |s|^{5/2} \right)\, .  
\ee
Note that in this edge regime quantum correlations
are still important. At much higher temperatures $T \sim N$, the positions
of the fermions become completely independent variables,
and the fluctuation of ${\sf x}_{\rm max}(T)$ is also described by a 
Gumbel distribution, albeit different from the one obtained here
\cite{longUS}.

The above method is easily extended to obtain the full counting statistics (FCS)
of the fermions near the edge for temperatures $T\gg N^{1/3}$. This is a generalisation of the
$T=0$ result for the FCS in the edge regime~\cite{Eisler2013}. Denoting by $N(s)$ 
the number of fermions in the interval $[{\sf x}_{\rm edge} + s\, w_N(T),+\infty[$ {we obtain
the characteristic function (see (\ref{FCS}) in \cite{SuppMat}) and, from it, the cumulants 
\bea
&& \langle \big( N(s) \big)^p  \rangle^c \simeq - \sqrt{\frac{T^3}{4 \pi N}}  \, Li_{\frac{5}{2}-p}( - e^{- s} ) 
\label{cum1} 
\eea 
for all positive integer $p \geq 1$.
In the typical region defined above, $s - \xi_{\rm typ} = {\cal O}(1)$ 
the
statistics is Poisson with mean
$\langle N(s) \rangle \simeq e^{\xi_{\rm typ}-s}$.}
There are deviations from Poisson in the
tails, in particular for $s - \xi_{\rm typ} \to - \infty$ where the distribution becomes
peaked around 
$\langle N(s) \rangle  \simeq \sqrt{\frac{T^3}{4 \pi N}} \frac{4 (-s)^{3/2}}{3 \sqrt{\pi }}$ 
with
$\langle N(s)^2 \rangle^c  \simeq \sqrt{\frac{T^3}{4 \pi N}} \frac{2 \sqrt{-s}}{\sqrt{\pi }}$
and zero higher cumulants. 



In conclusion we have studied the statistics of the height
fluctuations for the continuum KPZ equation at short time
with the droplet initial condition. We obtained the exact
analytical rate function $\Phi_{\rm drop}(H)$ and compared with
numerics. It confirms, and extends, through an exact solution, recent approaches
using weak noise theory developed for the flat geometry
and unveils puzzling similarities with other large deviation
results for finite-size system. 
We demonstrate that, remarkably, the right tail coincides with the Tracy-Widom
result already at short time. This result agrees with rigorous bounds \cite{Chen} 
valid at any fixed time $t$, $P(H > s) \leq e^{- \frac{4}{3} s^{3/2}/t^{1/2}}$. 
By contrast the convergence towards the
left TW tail $\sim (- H)^3$ appears to be much slower, with 
$\sim (-H)^{5/2}$ behavior at short time. Our short-time results for the KPZ equation 
also provide exact asymptotic predictions for 
the PDF of the rightmost fermion in a harmonic trap at high temperature $T\gg N^{1/3}$. We hope that
the present results will stimulate further investigations of extreme
value questions in the KPZ class~\cite{Khoshnevisan} and also in cold atom systems.

%

We thank D. S. Dean, K. Johansson, D. Khosnevisan, 
B. Meerson, J. Quastel, T. Sadhu, H. Spohn and
K. Takeuchi
for useful discussions. We acknowledge support from PSL grant ANR-10-IDEX-0001-02-PSL
(PLD). 
We thank the hospitality of KITP, under Grant No. NSF PHY11-25915.



{}

\newpage

.

\begin{widetext} 

\bigskip

\bigskip

\begin{large}
\begin{center}

SUPPLEMENTARY MATERIAL

\end{center}
\end{large}

\bigskip

We give the principal details of the calculations described in the manuscript of the Letter. 

\section{1. Short time estimate of the Fredholm determinant $Q_t(s)$}

We start by deriving the formula for $Q_t(s)$ given in Eq. (\ref{resQ}) in the Letter. From Eqs. (\ref{FD1}) and (\ref{sump}) given in the Letter, one has
\begin{eqnarray}\label{Q_start_supp}
\ln Q_t(s) = - \sum_{p=1}^\infty \frac{1}{p} {\rm Tr}\, \bar{K}^p_{t,s} \;, \; \bar K_{t,s}(u,u') = K_{\rm Ai}(u,u') \sigma_{t,s}(u')
\end{eqnarray}
where $K_{\rm Ai}(u,u')$, the Airy kernel, and $\sigma_{t,s}$ are given in Eqs. (\ref{defAiry1}) and (\ref{sig}) of the Letter (respectively). 
Hence one has
\bea
&& {\rm Tr} \; {\bar K}_{t,s}^p = \int_{-\infty}^\infty dv_1 \int_{-\infty}^\infty dv_2 \ldots \int_{-\infty}^\infty dv_p K_{\rm Ai}(v_1,v_2) .. K_{\rm Ai}(v_p,v_1)  \sigma_{t,s}(v_1) \ldots  \sigma_{t,s}(v_p)
\eea
The expression of $\sigma_{t,s}(v) = \sigma(t^{1/3}(v-s))$ suggests to perform the change of variable $v_i \to v_i / t^{1/3}$, which yields (setting $\tilde s = s t^{1/3}$):
 \bea\label{trace_Kp}
{\rm Tr} \; {\bar K}_{t,s}^p &=& t^{-p/3} \int_{-\infty}^\infty dv_1 \int_{-\infty}^\infty dv_2 \ldots \int_{-\infty}^\infty dv_p \, K_{\rm Ai}\left(\frac{v_1}{t^{1/3}} ,\frac{v_2}{t^{1/3}} \right) \ldots K_{\rm Ai}\left(\frac{v_p}{t^{1/3}} ,\frac{v_1}{t^{1/3}} \right)  
\sigma(v_1-\tilde s) \ldots \sigma(v_p-\tilde s) \\
\sigma(v) &=& \frac{1}{e^{- v } + 1} \;. \nonumber
\eea 
{Let us now recall the two useful representations of the Airy kernel
\be
K_{\rm Ai}(u,u') = \int_0^{+\infty} dr Ai(r+u) Ai(r+u') \label{defAiry1new}  = \frac{Ai(u) Ai'(u') - Ai(u') Ai'(u')}{u-u'} \;.
\ee
From the second expression
}, and using the asymptotic expansion of the Airy function for the large negative argument ${\rm Ai}(x) \sim \cos{(\pi/4 - 2|x|^{3/2}/3)}/\sqrt{\pi |x|^{1/2}}$, for $x \to -\infty$, one obtains the limiting form of the Airy kernel as
\bea
\lim_{t \to 0, v_1<0} t^{1/6} K_{\rm Ai}\left(\frac{v_1}{t^{1/3}} ,\frac{v_1 + t^{1/2} w}{t^{1/3}}\right)  = 
\frac{ 1}{\pi } \frac{\sin (\sqrt{|v_1|} w)}{w}  \;.
\eea 
On the other hand, for $v_1 > 0$, the Airy kernel vanishes exponentially in the limit $t \to 0$ 
and therefore only the region where all the $v_i$ are negative need to be considered
in Eq. (\ref{trace_Kp}). Hence for $p \geq 2$, 
separating the center of mass coordinate (which we take as $v_1$) 
and the $p-1$ relative coordinates $v_j=v_{j-1}+ t^{1/2} w_j$ 
we obtain
\bea
&&{\rm Tr} \bar{K}^p \simeq t^{-p/3}   \int_{-\infty}^0 dv_1 \left(\frac{1}{\pi t^{1/6}}\right)^p [\sigma(v_1-\tilde s)]^p 
t^{(p-1)/2}  \\
&& \times \int_{-\infty}^\infty dw_1  \ldots \int_{-\infty}^\infty dw_p  \frac{\sin ( \sqrt{|v_1|} w_1)}{w_1} \frac{\sin ( \sqrt{|v_1|} w_2)}{w_2} \ldots \frac{\sin ( \sqrt{|v_1|} w_p)}{w_p}
\delta(w_1+w_2+..+w_p) \nonumber
\\
&& = \frac{1}{\pi^p \sqrt{t} }  \int_{-\infty}^0 dv_1 \sqrt{|v_1|}   [\sigma(v_1-\tilde s)]^p I_p \;, \; I_p = 
\int_{-\infty}^\infty dw_1  \ldots \int_{-\infty}^\infty dw_p  \frac{\sin w_1}{w_1} \frac{\sin w_2}{w_2} .. \frac{\sin w_p}{w_p} \delta(w_1+..+w_p) \;. \label{Eq:Ip}
\eea
The multiple integral defining $I_p$ in Eq. (\ref{Eq:Ip}) can be computed explicitly, using $\sin{x}/x = (1/2) \int_{-1}^1 e^{i k x} dk$ and an integral representation of the delta function in Eq. (\ref{Eq:Ip}), to obtain
\bea
I_p= \frac{1}{2^p} \int_{-1}^1 dx_1 \ldots   \int_{-1}^1 dx_p \int_{-\infty}^\infty \frac{dk}{2 \pi} 
\int_{-\infty}^\infty dw_1 \ldots \int_{-\infty}^\infty dw_p e^{i \sum_{j=1}^p (x_j + k) w_j } = \frac{1}{2^p} (2 \pi)^p \int_{-1}^1  \frac{dk}{2 \pi}  =
\pi^{p-1} \;. \label{Eq:_i_final}
\eea 
Thus, from Eq. (\ref{Q_start_supp}) together with Eqs. (\ref{Eq:Ip}) and (\ref{Eq:_i_final}), one obtains 
\bea
&& \ln Q_t(s) \approx -\frac{1}{\sqrt{t}} \Psi(e^{-\tilde s}) \\
&& \Psi(z) = \frac{1}{\pi} \sum_{p=1}^\infty \frac{1}{p}  \int_{-\infty}^0 dv \sqrt{|v|} \frac{1}{(e^{- v } z^{-1}+ 1)^p}  = \frac{1}{\pi}
\sum_{p=1}^\infty \frac{1}{p}  \int_{0}^{+\infty} dv \sqrt{v}
\left(\frac{z e^{-v}}{(1+ z e^{-v})}\right)^p
\eea 
It is then straightforward to perform the sum over $p$ to get
\bea
\Psi(z) &=&  - \frac{1}{\pi} \int_{0}^{+\infty} dv \sqrt{v} \ln\left(1 - \frac{z e^{-v}}{1+ z e^{-v}}\right) 
= \frac{1}{\pi} \int_{0}^{+\infty} dv \sqrt{v} \ln(1+ z e^{-v}) \\
&=& - \frac{1}{\sqrt{4 \pi}} Li_{5/2}(-z) \;,
\eea 
as given in the Letter in Eq. (\ref{resQ}). 

\section{2. Counting statistics from a generalized Fredholm determinant} 

{We give here the details of the calculation of the characteristic function for the full counting statistics,
and from it the cumulants of the fermion number (\ref{cum1}) displayed in the text,} since it is a very simple modification
of the previous calculation. We use the fact that the quantum probability measure on the
fermion positions $x_i$ becomes a determinantal process in the limit
of large $N$ \cite{FermionsUS_app}.

Denoting as in the text, $N(s)$, the total number of fermions in the
interval $[s,+\infty[$ we can use the standard property of a determinantal
process to express the Laplace transform of its distribution as
\bea
\langle e^{- \lambda N(s) } \rangle = {\rm Det}[I - (1- e^{- \lambda}) P_0 K_{t,s} P_0 ] 
= {\rm Det}[I - (1- e^{- \lambda}) \bar K_{t,s}  ] 
\eea 
where $K_{t,s}$ and $\bar K_{t,s}$ are defined respectively in (\ref{Kt}) and (\ref{defKbar}) in the text. 
So it is a simple generalization
of $Q_t(s)$, which is recovered for $\lambda=+\infty$. It can thus also be 
expanded in traces of powers
\bea
\ln \langle e^{- \lambda N(s) } \rangle =  - \sum_{p=1}^{+\infty} \frac{1}{p}
(1- e^{- \lambda})^p  \, {\rm Tr}\,  \bar K_{t,s}^p
\eea 
Following the same steps as in the previous section, we thus obtain
\bea
&& \ln \langle e^{- \lambda N(s) } \rangle \approx -\frac{1}{\sqrt{t}} \Psi(e^{-\tilde s},\lambda) \\
&&  \Psi(z,\lambda)  = \frac{1}{\pi}
\sum_{p=1}^\infty \frac{1}{p}  \int_{0}^{+\infty} dv \sqrt{v}
\left(\frac{z (1- e^{- \lambda}) e^{-v}}{(1+ z e^{-v})}\right)^p =
- \frac{1}{\pi} \int_{0}^{+\infty} dv \sqrt{v} \ln\left(1 - \frac{z (1- e^{- \lambda})  e^{-v}}{1+ z e^{-v}}\right) \\
&& 
= \frac{1}{\pi} \int_{0}^{+\infty} dv \sqrt{v} ( \ln(1+ z e^{-v}) -  \ln(1+ z e^{- \lambda} e^{-v}) ) = 
- \frac{1}{\sqrt{4 \pi}} ( Li_{5/2}(-z)  - Li_{5/2}(-z e^{- \lambda}) )
\eea  
In the notations of the text, $t \to b^3=N/T^3$ and $z \to e^s$ (note that $\tilde s$ is denoted simply $s$ in
the part on the fermions) {we obtain 
the characteristic function (as the Laplace transform) 
\be
\ln \langle e^{- \lambda N(s) } \rangle \simeq  \sqrt{\frac{T^3}{4 \pi N}} 
[ Li_{\frac{5}{2}}( - e^{- s}) - Li_{\frac{5}{2}}( - e^{- s - \lambda}) ]  \label{FCS} 
\ee 
from which the cumulants (\ref{cum1}) given in the text are easily extracted.
Note that the fact that in the typical region $s - \xi_{\rm typ} = {\cal O}(1)$, the statistics is Poisson
can also be seen directly on the above characteristic function since in that limit
$\ln \langle e^{- \lambda N(s) } \rangle \simeq - e^{\xi_{\rm typ}-s} (1-e^{-\lambda})$.}

\section{3. Evaluation of $\Phi_{\rm drop}(H)$}

We start from Eq. (\ref{genfunct}) of the text that reads
\begin{equation}
\left\langle \exp\left(-\frac{z}{\sqrt{4\pi t}}\, 
e^{H(t)}\right)\right\rangle \sim 
e^{-\frac{1}{\sqrt{t}}\, \Psi(z)}
\label{supp1.1}
\end{equation}
where $\Psi(z)= -Li_{5/2}(-z)/\sqrt{4\pi}$ is given in Eq. (25) of the 
text. Substituting the anticipated form, $P(H,t) \sim 
e^{-\frac{1}{\sqrt{t}}\, \Phi(H)}$ as $t\to 0$, on the lhs of 
Eq. (\ref{supp1.1})
gives
\begin{equation}
\left\langle \exp\left(-\frac{z}{\sqrt{4\pi t}}\, 
e^{H(t)}\right)\right\rangle\sim
\int dH\, \exp\left[-\frac{1}{\sqrt t}\, \left(\frac{z}{\sqrt{4\pi}}\, e^H 
+ \Phi_{\rm drop}(H)\right)\right]\, .
\label{supp1.2}
\end{equation}
Using $1/\sqrt{t}$ as a large parameter as $t\to 0$, the integral can be
evaluated by the saddle point and comparing it to the rhs of Eq. 
(\ref{supp1.1}) gives 
\begin{equation}
\min_{H}\left[ \frac{z}{\sqrt{4\pi}}\, e^H + \Phi_{\rm drop}(H)\right]= \Psi(z)\, .
\label{supp1.3}
\end{equation}
Inverting this Legendre transform (assuming convexity of $\Phi(H)$) one 
gets
\begin{eqnarray}
\Phi_{\rm drop}(H)&= &\max_{z}\left[- \frac{z}{\sqrt{4\pi}}\,e^H + \Psi(z)\right] 
\label{supp1.4} \\
&= & -\frac{1}{\sqrt{4\pi}}\min_{z}\left[ z\, e^H + Li_{5/2}(-z)\right]\, ,
\label{supp1.5}
\end{eqnarray}
where we used $\Psi(z)= -Li_{5/2}(-z)/\sqrt{4\pi}$.
Deriving $S_1(z)\equiv z e^H + Li_{5/2}(-z)$ with respect to $z$ 
determines
the minimizer $z^*$, for a given $H$, as 
\begin{equation}
e^H= -\frac{1}{z^*}\, Li_{3/2}(-z^*)\equiv W_1(z^*)\, ,
\label{supp1.6}
\end{equation}
where we used $\frac{d}{dz} Li_{5/2}(-z)= \frac{1}{z}Li_{3/2}(-z)$.
The function $W_1(z^*)$ in Eq. (\ref{supp1.6}) is convergent only
in the range $z^*\in [-1,\infty]$ and has the 
following asymptotic properties
\begin{eqnarray}
W_1(z^*) &\simeq & -\frac{1}{\Gamma(5/2)}\, 
\frac{\left(\ln z^*\right))^{3/2}}{z^*} 
\quad\,\, {\rm 
as}\quad z^*\to \infty \label{supp1.7} \\
& \simeq & \zeta(3/2)=2.61238\dots \quad {\rm as}\quad z^*\to -1 \, .
\label{supp1.8}
\end{eqnarray}
As one decreases $z^*$ from $\infty$ to $-1$, $W_1(z^*)$ thus increases
monotonically from $0$ to $\zeta(3/2)=2.61238\dots$ (shown by the 
solid (black) line in Fig.(\ref{fig.wz})). Thus, for 
any given $H\in \left[-\infty, H_c=\ln(\zeta(3/2)=0.96026\dots\right]$, 
there is
a unique solution $z^*(H)$ of Eq. (\ref{supp1.6}). 

Naturally, the question arises: how do we find a solution for $H>H_c$?
Interestingly, a similar minimization problem also appeared in
the compuation of the large deviation function in the asymmetric exclusion 
problem in a finite ring~\cite{DL1998,DA1999}. The trick is to use
the analytically continued partner of $Li_{5/2}(-z)$ (instead of 
$Li_{5/2}(-z)$) on the rhs of Eq. (\ref{supp1.5}). The correct
analytically continued partner~\cite{DL1998,DA1999} of $Li_{5/2}(-z)$ 
turns out to 
be, 
$Li_{5/2}(-z)- \frac{8\sqrt{\pi}}{3}\left[-\ln(-z)\right]^{3/2}$ where $z$ 
now
increases back from $-1$ to $0$. In other
words, Eq. (\ref{supp1.5}) is now replaced (for $H>H_c$) by
\begin{equation}
\Phi_{\rm drop}(H)= -\frac{1}{\sqrt{4\pi}}\min_{z}\left[ z\, e^H + Li_{5/2}(-z) -
\frac{8\sqrt{\pi}}{3}\, \left[-\ln(-z)\right]^{3/2}\right]\, .
\label{supp1.9}
\end{equation}
Deriving $S_2(z)\equiv z\, e^H + Li_{5/2}(-z) -
\frac{8\sqrt{\pi}}{3}\, \left[-\ln(-z)\right]^{3/2}$ with respect to $z$ 
now provides  
the minimizer $z^*$ for $H>H_c$
\begin{equation}
e^H= -\frac{1}{z^*}\, Li_{3/2}(-z^*) 
-\frac{4\sqrt{\pi}}{z^*}\left[-\ln(-z^*)\right]^{1/2}\equiv W_2(z^*) \, .
\label{supp1.10}
\end{equation}
The function $W_2(z^*)$ is defined for all $z^*\in [-1,0]$.
As $z^*$ increases from $-1$ to $0$,
the function $W_2(z^*)$ in Eq. (\ref{supp1.10}) increases monotonically
(shown by the dashed (red) line in Fig. (\ref{fig.wz})), with the 
following 
limiting behaviors 
\begin{eqnarray}
W_2(z^*) & =& \zeta(3/2)=2.61238\dots \quad {\rm as}\quad z^*\to -1 
\label{supp1.11} \\
& \simeq & -\frac{4\sqrt{\pi}}{z^*}\left[-\ln(-z^*)\right]^{1/2} \quad 
{\rm 
as}\quad z^*\to 0^{-}\, . \label{supp1.12}
\end{eqnarray}
Thus, in this range, one can find a unique solution $z^*(H)$ of Eq. 
(\ref{supp1.10}) for any $H\in [H_c=\ln(\zeta(3/2)), \infty]$.
\begin{figure}
\begin{center}
\includegraphics[width = 0.6\linewidth]{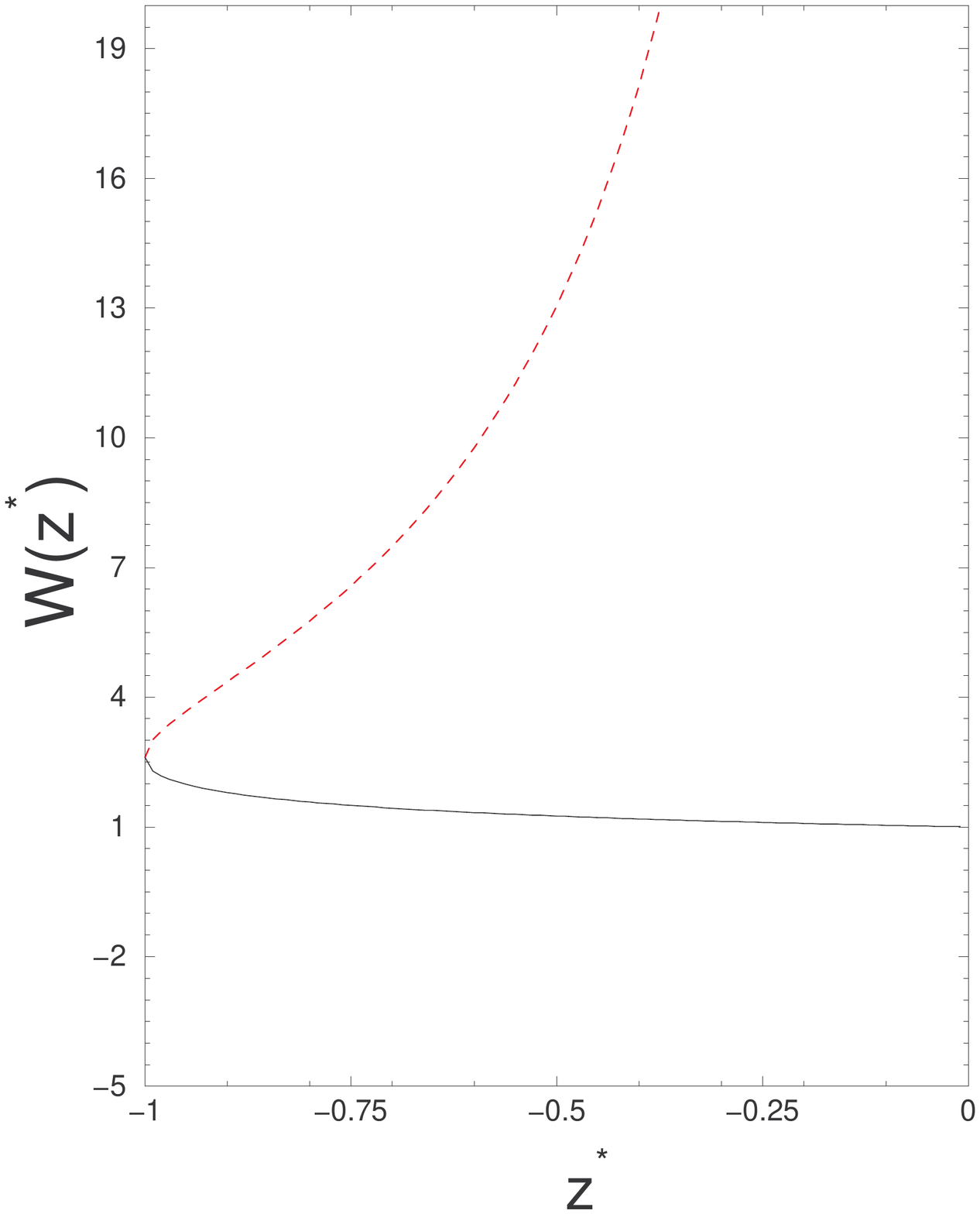}
\caption{The function $W(z^*)$ vs $z^*$ plotted in the range $z^*\in 
[-1,0]$. The branch $W_1(z^*)$ with
$z^*\in [-1,\infty]$ is shown by the solid (black) line (plotted
only in the regine $z^*\in [-1,0]$ for convenience). The 
branch
$W_2(z^*)$ with $z^*\in [-1,0]$ is shown by the dashed (red) line.
They join smoothly at $z^*=-1$ where $W(z^*)= \zeta(3/2)=2.61238\dots$.} 
\label{fig.wz}
\end{center}
\end{figure}

To summarize, for a given $H\in [-\infty,\infty]$, the minimizer
$z^*$ is determined from the equation
\begin{equation}
e^H= W(z^*)\, 
\label{supp1.13}
\end{equation}
where the function $W(z^*)$ is given by
\begin{eqnarray}
W(z^*)&= & W_1(z^*)= -\frac{1}{z^*}\, Li_{3/2}(-z^*)\quad {\rm 
for}\quad 
z^*\in [-1,\infty]\quad {\rm and}\quad H<H_c \label{supp1.14}\\
W(z^*) &= & W_2(z^*)= -\frac{1}{z^*}\, Li_{3/2}(-z^*)
-\frac{4\sqrt{\pi}}{z^*}\left[-\ln(-z^*)\right]^{1/2} \quad {\rm for}\quad
z^*\in [-1,0]\quad {\rm and}\quad H>H_c \, . \label{supp1.15}
\end{eqnarray}
The function $W(z^*)$ vs $z^*$ is plotted in Fig. (\ref{fig.wz}), with
the two branches $W_1(z^*)$ (shown by solid (black) line) and $W_2(z^*)$
(shown by the dashed (red) line). We remark that there is no phase
transition at $H=H_c$, as the function $\Phi(H)$ is analytic at $H=H_c$. 

\section{4. Asymptotic behavior of $\Phi_{\rm drop}(H)$} 

Let us derive here the left tail behavior of $\Phi_{\rm drop}(H)$, for $H \to - \infty$.
From the saddle point equation (\ref{supp1.6}) we see that it corresponds
to $z^* \to +\infty$. In that limit we can use the following estimate 
and $\nu \in \mathbb{N}/2$, $\nu \geq 3/2$ \cite{Polylog}:
\bea
Li_\nu(-z) \simeq_{z \to +\infty} - \frac{1}{\Gamma(1+ \nu)} (\ln z)^{\nu} - \frac{\pi^2}{6 \Gamma(\nu-1)} (\ln z)^{\nu-2} + .. 
\eea 
Hence for $z \to +\infty$ we find that
\bea
&& H \simeq - \ln z + \ln ( \frac{4}{3 \sqrt{\pi}} (\ln z)^{3/2} ) + ..
\eea 
Inserting back in the formula  (\ref{supp1.5}) we finally
find
\bea
&& \Phi_{\rm drop}(H) \simeq  \frac{4}{15 \pi}  (-H)^{5/2} + \frac{1}{\pi} (-H)^{3/2} 
\left(\ln (-H) + \frac{2}{3}\left( \ln(\frac{4}{3 \sqrt{\pi}})-1\right) \right) + .. 
\eea 
which is given in the text.

To obtain the right tail of $\Phi_{\rm drop}(H)$ we write the saddle point equation
(\ref{supp1.10}). For large $H$ which corresponds to $z^* \to 0$ 
we can use the asymptotic behavior in (\ref{supp1.12}),
i.e. $H \simeq - \ln (-z^*) + \frac{1}{2} \ln H + \ln(4 \sqrt{\pi})$. Reinserting this value of $H$ in (\ref{supp1.9})
evaluated at $z=z^*$ we find the two leading orders
\bea
&& \Phi_{\rm drop}(H) \simeq_{H \to +\infty} \frac{4}{3} H^{3/2} - ( \ln H + 2 \ln(4 \sqrt{\pi}) -2) H^{1/2} + .. 
\eea 
which is given in the text.

\section{5. Short time cumulants of $H$ and relation to 
Derrida-Lebowitz cumulants}

To compute the cumulants of the height $H(t)$ at short times, we first
define the cumulant generating function
\begin{equation}
G(p,t)= \left\langle e^{\frac{p}{\sqrt{t}}\, H}\right\rangle= \int 
e^{\frac{p}{\sqrt{t}}\,
H}\, 
P(H,t)\, dH\, ,
\label{cum1.1}
\end{equation}
where $P(H,t)$ is the height pdf. Substituting the short time form,
$P(H,t)\sim e^{-\frac{1}{\sqrt{t}}\, \Phi_{\rm drop}(H)}$ in Eq. (\ref{cum1.1})
and performing the integral by the saddle point method as $t\to 0$ gives
\begin{equation}
G(p,t)\simeq e^{\frac{1}{\sqrt{t}}\, \phi(p)},\quad {\rm where}\quad 
\phi(p)= \max_{H}\left[p H - \Phi_{\rm drop}(H)\right]\, ,
\label{cum1.2}
\end{equation}
where $\Phi_{\rm drop}(H)$ is given explicitly in Eq. (27) of the text. We note
that, be definition, the logarithm of $G(p,t)$ generates the height 
cumulants by the 
\begin{equation}
\ln G(p,t)= \sum_{q=1}^{\infty} \overline{H(t)^q}^c\, 
\left[\frac{p}{\sqrt{t}}\right]^q \, .
\label{cum1.3}
\end{equation}
Hence, taking logarithm on both sides of Eq. (\ref{cum1.2}), using
(cum1.3) and matching powers of $p$ gives
\begin{equation}
\overline{H(t)^q}^c= t^{(q-1)/2}\, \phi^{(q)}(0)\, ,
\label{cum1.4}
\end{equation}
for all $q\ge 1$,
where $\phi^{(q)}(0)$ is the $q$-th derivative of $\phi(p)=
\max_{H}\left[p H - \Phi_{\rm drop}(H)\right]$ evaluated at $p=0$. Note that the 
centering of $H$ makes the
first cumulant vanish.
Using the
explicit form of $\Phi_{\rm drop}(H)$ in Eq. (27) of the text, one can
obtain $\phi^{(q)}(0)$ explicitly. For example, the first $3$ nonzero
cumulants are given by
\begin{eqnarray}
\phi^{(2)}(0) &= & \sqrt{\frac{\pi}{2}} \\
\phi^{(3)}(0) & = & \left(\frac{8}{3\sqrt{3}}-\frac{3}{2}\right)\pi \\
\phi^{(4)}(0) & = & 
\frac{1}{3}\,\left(18+15\sqrt{2}-16\sqrt{6}\right)\pi^{3/2} \,  \\
\phi^{(5)}(0) & = & 
120\,\left(-\frac{317}{432}-\frac{1}{\sqrt{2}}+\frac{2}{\sqrt{3}}+
\frac{16}{25\sqrt{5}}\right) \pi^2 
\label{cum1.5}
\end{eqnarray} 

Remarkably, these cumulants carry an uncanny resemblance to the
late time cumulants of the total integrated current in the totally 
asymmetric exclusion process (TASEP) on a ring of size $N$, derived
by Derrida and Lebowitz~\cite{DL1998}. 
More 
precisely, Derrida and Lebowitz considered the TASEP on a finite ring of 
size $N$ with a fixed density $\rho$ of hard core particles. 
Each particle attempts a jump to the neighboring site with rate $1$
and succeeds provided the target site is empty. Let $J_i(T)$ denote
the total current up to time $t$ through the bond $i$, i.e., the total
number of particles that have passed through the bond $i$ up to time $T$.
They considered the random variable $Y_T= \sum_{i=1}^N J_i(T)$ denoting
the total integrated current in the system up to time $T$. Using Bethe
ansatz techniques, they were able to compute exactly the cumulants
of $Y_T$ at late times $ T>> N^{3/2}$. The first three nonzero moments
are given by~\cite{DL1998}
\begin{eqnarray}
\overline{Y_T^2}^c &=& T\, N^{3/2}\, 
\left[\rho(1-\rho)\right]^{3/2}\, \frac{\sqrt{\pi}}{2} \\
\overline{Y_T^3}^c &=& T\, N^3\, \left[\rho(1-\rho)\right]^{2}\,
\left(\frac{3}{2}-\frac{8}{3 \sqrt{3}}\right)\, \pi \\
\overline{Y_T^4}^c &=& T\, N^{9/2}\, \left[\rho(1-\rho)\right]^{5/2}\,
(18+15\sqrt{2}-16 \sqrt{6}) \frac{\pi^{3/2}}{2 \sqrt{2}} 
\\
\overline{Y_T^5}^c &=& T\, N^{6}\, \left[\rho(1-\rho)\right]^{3}\,
120\,\left(-\frac{317}{432}-\frac{1}{\sqrt{2}}+\frac{2}{\sqrt{3}}+
\frac{16}{25\sqrt{5}}\right)
\pi^2 \, .
\label{cum1.6}
\end{eqnarray}
Naively, the numerical factors on the rhs of Eq. (\ref{cum1.6}), 
do not look similar to the numerical factors on the rhs of Eq. 
(\ref{cum1.5}). Remarkably, when slightly re-arranged, they however look
very similar! To see this more clearly, we define
the ratio
\begin{equation}
R_q^c= \frac{\overline{Y_T^q}^c}{\overline{H(t)_t^q}^c} \,.
\label{cum1.7}
\end{equation}
From Eqs. (\ref{cum1.5}) and (\ref{cum1.6}), one finds that the ratio
is rather simple and for general $2\le q\le 5$, it reads
\begin{equation}
R_q^c= (-1)^q (q-1)\, \, \frac{T}{(2t)^{(q-1)/2}}\, N^{3(q-1)/2}\, 
\left[\rho(1-\rho)\right]^{(q+1)/2}\, , 
\label{cum1.8}
\end{equation} 
with all the strange looking numerical factors disappearing totally!
By computing higher cumulants (not shown here) in the two problems,
we have verified that this relation holds also for all integer $q>5$.

We do not quite understand why the numerical factors in the 
cumulants of these two problems are so simply related. First of all,
in the TASEP problem one is considering at a finite size ($N$) system
and at {\em late} times $T>> N^{3/2}$. Using a mapping between 
TASEP and a discrete growth model, $Y_T$ would translate into the
integrated height $Y_T\equiv \sum_{i=1}^N H_i(T)$ where
$H_i(T)$ denotes the height at site $i$ of the discrete-time growth
model~\cite{DA1999}. Still, the results of Derrida nd Lebowitz hold
only at  
{\em late} times $T>>N^{3/2}$.
In contrast, in the continuum KPZ equation studied in this paper,
we compute the cumulants of the height 
$H(0,t)$, but at {\em short} times $t\to 0$ in an already infinite 
system. So, 
even admitting the universality of the KPZ growth equation, there 
is no apriori reason why these two observables in very different 
time regimes should have a simple relation between their cumulants.
This remains an outstanding puzzle to be understood fully.

{

\section{6. Directed polymer model and numerical details}

To test the validity of our results numerically we simulate a directed polymer (DP) growing on a two dimensional square lattice. We define the partition sum $\tilde Z_{i,j}=\sum_{\gamma} e^{- \beta \sum_{(r,s) \in \gamma} V_{r,s}}$
over all paths $\gamma$ directed along the diagonal on a square lattice, with only $(1,0)$ or $(0,1)$ moves, starting in $(0,0)$ and 
ending in $(i,j)$, where the $V_{r,s}$ are i.i.d. random site variables distributed with 
a unit centered Gaussian. Introducing ``time" $\hat t=i+j$ and space
$\hat x = \frac{i-j}{2}$, $Z_{\hat x,\hat t}=\tilde Z_{i,j}$ satisfies:
\be \label{discrete}
Z_{\hat x,\hat t+1}=(Z_{\hat x-\frac{1}{2},\hat t}+Z_{\hat x+ \frac{1}{2},\hat t}) e^{-\beta V_{\hat x,\hat t+1}}
\ee
with 
$Z_{\hat x,0}=\delta_{\hat x,0}$. 
As discussed in \cite{CLR10_app} the high temperature limit this DP model maps onto the continuum equation (\ref{eq:KPZ}) 
in terms of the variables $x=4 \hat x \beta^2$ and
$t=2 \hat t \beta^4$. The corresponding continuum height field 
(\ref{H}) at $x=0$ is obtained as $H \equiv \ln Z_{0, \hat t}/\langle Z_{0, \hat t} \rangle$.
The result for $P(H,t)$ is shown in the Fig. \ref{fig:PHIdrop2}.

}

{}

\end{widetext}


\begin{thebibliography}{}
\bibitem{footnote1}
Note that in our units $\Phi_{\rm flat}(H)=\frac{1}{8} S(-2 H)$ where $S(H)$ 
is given in \cite{Baruch}.
\bibitem{footnoteGaussian} 
Note that the coefficient of the central
Gaussian part depends on the initial condition.
\bibitem{footnote2}
The continuum solution is related to the physical solution up to a non-universal
shift (i.e. renormalization). Let us call $h^{\rm phys}(x,t)$ the solution to 
a physical KPZ problem with a regularized, i.e. 
smooth noise at small scale. Then, above a correspondingly small time and length scale,
the continuum and physical solutions are related as follows
$h(x,t)= \ln Z^{\rm phys}(x,t)/\langle Z^{\rm phys}(x,t) \rangle + \ln Z_0(x,t) =
h^{\rm phys}(x,t) - \ln( \langle \exp(h^{\rm phys}(x,t)) \rangle ) + \ln Z_0(x,t)$.
It holds for an arbitrary initial condition provided $Z_0(x,t)$ is chosen as the solution of
the free diffusion equation with that initial condition. See \cite{flatshorttime}
for a more detailed discussion in case of flat initial conditions.
\bibitem{footnote_Fredholm}
We recall that, for a trace-class operator $K(x,y)$ such that ${\rm Tr} K = \int dx K(x,x)$ is well 
defined, $\det(I - K) = \exp{[-\sum_{n=1}^\infty{{\rm Tr \,} K^n}/{n}]}$, where ${\rm Tr}\, K^n = 
\int dx_1 \cdots \int dx_n K(x_1,x_2) K(x_2,x_3)\cdots K(x_n,x_1)$. The effect of the projector 
$P_0$ in (\ref{FD1}) is simply to restrict the integrals over $x_i$'s to the interval $[0,+\infty 
)$.
\end{thebibliography}
\end{document}